\documentclass[aps,pre,twocolumn,superscriptaddress,showpacs]{revtex4-1}

\usepackage{color}
\usepackage{strike}

\usepackage{graphicx}
\usepackage{amsmath}

\begin{document}

\title{Interlayer-Interaction Dependence of Latent Heat in the Heisenberg Model \\ on a Stacked Triangular Lattice with Competing Interactions}

\author{Ryo Tamura}
\email[]{tamura.ryo@nims.go.jp}
\affiliation{International Center for Young Scientists, National Institute for Materials Science, 1-2-1, Sengen, Tsukuba-shi, Ibaraki, 305-0047, Japan}

\author{Shu Tanaka}
\email[]{shu-t@chem.s.u-tokyo.ac.jp}
\affiliation{Department of Chemistry, University of Tokyo, 7-3-1, Hongo, Bunkyo-ku, Tokyo, 113-0033, Japan}

 
\begin{abstract}
We study the phase transition behavior of a frustrated Heisenberg model on a stacked triangular lattice by Monte Carlo simulations.
The model has three types of interactions: the ferromagnetic nearest-neighbor interaction $J_1$ and antiferromagnetic third nearest-neighbor interaction $J_3$ in each triangular layer and the ferromagnetic interlayer interaction $J_\perp$.
Frustration comes from the intralayer interactions $J_1$ and $J_3$.
We focus on the case that the order parameter space is SO(3)$\times C_3$.
We find that the model exhibits a first-order phase transition with breaking of the SO(3) and $C_3$ symmetries at finite temperature.
We also discover that the transition temperature increases but the latent heat decreases as $J_\perp/J_1$ increases, which is opposite to the behavior observed in typical unfrustrated three-dimensional systems.
\end{abstract}



\maketitle


\section{Introduction}

\begin{figure}[b]
\begin{center}
\includegraphics[scale=1.0]{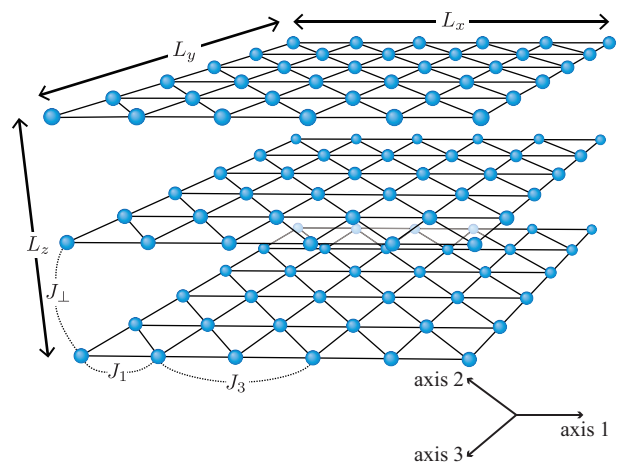} 
\end{center}
\caption{\label{fig:lattice}
(Color online) Schematic picture of a stacked triangular lattice with $L_x \times L_y \times L_z$ sites.
Here $J_1$ and $J_3$ respectively represent the nearest-neighbor and third-nearest-neighbor interactions in each triangular layer and $J_\perp$ is the interlayer interaction.
}
\end{figure}

Geometrically frustrated systems often exhibit a characteristic phase transition, such as successive phase transitions, order by disorder, and a reentrant phase transition, and an unconventional ground state, such as the spin liquid state \cite{Toulouse-1977,Liebmann-1986,Kawamura-1998,Diep-2005,Southern-2012,Villain-1980,Henley-1989,Fradkin-1976,Miyashita-1983,Kitatani-1985,Kitatani-1986,Azaria-1987,Miyashita-1987,Tanaka-2005,Tanaka-2010,Tanaka-2007,Ishii-2011,Shimizu-2003,Nakatsuji-2005,Balents-2010}.
In frustrated continuous spin systems, the ground state is often a noncollinear spiral-spin structure \cite{Yoshimori-1959,Nagamiya-1967}.
The spiral spin structure leads to exotic electronic properties such as multiferroic phenomena \cite{Katsura-2005,Kimura-2006,Cheong-2007,Tokura-2010,Arima-2011}, the anomalous Hall effect \cite{Nagaosa-2010}, and localization of electronic wave functions \cite{Tanaka-2006}.
Thus, the properties of frustrated systems have attracted attention in statistical physics and condensed matter physics.
Many geometrically frustrated systems such as stacked triangular antiferromagnets (see Fig.~\ref{fig:lattice}), stacked kagome antiferromagnets, and spin-ice systems have been synthesized and their properties have been investigated. 
In theoretical studies, the relation between phase transition and order parameter space in geometrically frustrated systems has been considered \cite{Miyashita-1984,Kawamura-1984,Loison-inBook-Diep,Okubo-2012,Tamura-2013}.

As an example of phase transition nature in geometrically frustrated systems, properties of the Heisenberg model on a triangular lattice have been theoretically studied for a long time.
Triangular antiferromagnetic systems are a typical example of geometrically frustrated systems and have been well investigated.
The ground state of the ferromagnetic Heisenberg model on a triangular lattice is a ferromagnetically collinear spin structure.
In this case, the order parameter space is $S_2$.
The long-range order of spins does not appear at finite temperature because of the Mermin-Wagner theorem \cite{Mermin-1966}.
The model does not exhibit any phase transitions.
In contrast, Refs.~\cite{Kawamura-1984,Kawamura-2010,Kawamura-2011} 
reported that a topological phase transition occurs in the Heisenberg model on a triangular lattice with only antiferromagnetic nearest-neighbor interactions.
In this model the long-range order of spins is prohibited by the Mermin-Wagner theorem, and thus a phase transition driven by the long-range order of spins never occurs as well as in the ferromagnetic Heisenberg model.
Since the ground state of the model is the $120^\circ$ structure, the order parameter space is SO(3), which is the global rotational symmetry of spins.
Thus the point defect, i.e., $Z_2=\pi_1 ({\rm SO(3)})$ vortex defect, can exist in the model.
Then the topological phase transition occurs by dissociating the $Z_2$ vortices at finite temperature \cite{Kawamura-1984,Kawamura-2010,Kawamura-2011}.
The dissociation of $Z_2$ vortices is one of the characteristic properties of geometrically frustrated systems when the ground state is a noncollinear spin structure in two dimensions.
In these systems, the order parameter space is described by SO(3).
The temperature dependence of the vector chirality and that of the number density of $Z_2$ vortices in the Heisenberg model on a kagome lattice were also studied \cite{Domenge-2008}.
An indication of the $Z_2$ vortex dissociation has been observed in electron paramagnetic resonance (EPR) and electron spin resonance (ESR) measurements \cite{Ajiro-1988,Yamaguchi-2008,Hemmida-2011}.

Phase transition has been studied theoretically in stacked triangular lattice systems as well as in two-dimensional triangular lattice systems.
In many cases, the phase transition nature in three-dimensional systems differs from that in two-dimensional systems.
In the Heisenberg model on a stacked triangular lattice with the antiferromagnetic nearest-neighbor intralayer interaction $J_1$ and the nearest-neighbor interlayer interaction $J_\perp$, the ground state is a $120^\circ$ structure in each triangular layer.
Thus, the order parameter space is SO(3) as in the two-dimensional case.
Two types of contradictory results have been reported.
In one, a second-order phase transition belonging to the universality class called the chiral universality class, which relates to the SO(3) symmetry, occurs \cite{Kawamura-1985,Kawamura-1992,Kawamura-1998,Pelissetto-2001,Calabrese-2004,Murtazaev-2007}.
In the other, a first-order phase transition occurs at finite temperature \cite{Zumbach-1993,Tissier-2000,Zelli-2007,Ngo-2008}.
In either case, the phase transition nature in the Heisenberg model on a stacked triangular lattice differs from that on a two-dimensional triangular lattice.

Recently, another kind of characteristic phase transition nature has been found in Heisenberg models on a triangular lattice with further interactions \cite{Tamura-2008,Tamura-2011,Stoudenmire-2009,Okubo-2012,Tamura-2013}.
The order parameter space is described by the direct product between the global rotational symmetry of spins SO(3) and discrete lattice rotational symmetry, which depends on the ground state.
In these models, a phase transition with the discrete symmetry breaking occurs at finite temperature.
In the $J_1$-$J_3$ Heisenberg model on a triangular lattice, 
the ground state is the spiral-spin structure where $C_3$ lattice rotational symmetry is broken due to the competition between the ferromagnetic nearest-neighbor interaction $J_1$ and antiferromagnetic third nearest-neighbor interaction $J_3$ \cite{Tamura-2008,Tamura-2011}.
In this case, the order parameter space is SO(3)$\times C_3$.
This model exhibits a first-order phase transition with breaking of the $C_3$ symmetry.
In addition, the dissociation of $Z_2$ vortices that comes from the SO(3) symmetry occurs at the first-order phase transition temperature.
A similar phase transition with the discrete symmetry breaking also has been found in Heisenberg models on square and hexagonal lattices with further interactions \cite{Loison-2000,Weber-2003,Okumura-2010}.
To consider a microscopic mechanism of the first-order phase transition with the discrete symmetry breaking in frustrated continuous spin systems, 
a generalized Potts model, called the Potts model with invisible states, has been studied \cite{Tamura-2010,Tanaka-2011a,Tanaka-2011b}.

\begin{figure}[t]
\begin{center}
\includegraphics[scale=1.0]{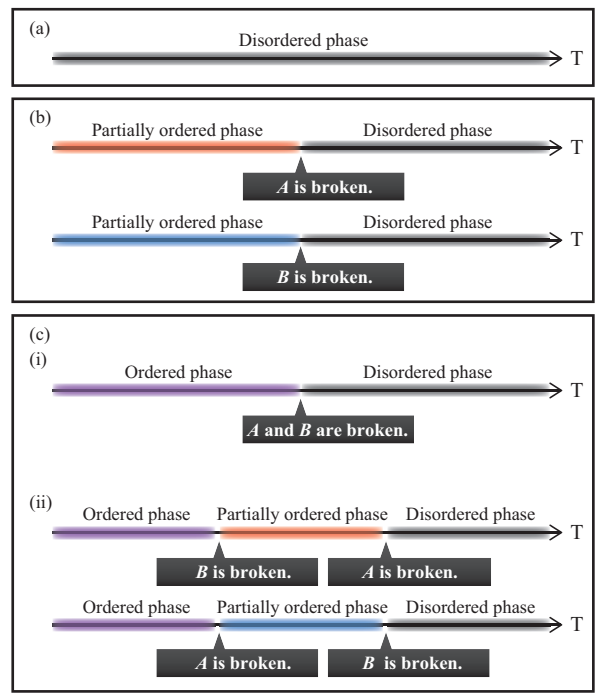} 
\end{center}
\caption{\label{fig:phasetransition_concept}
(Color online) Schematic of the phase transition nature in systems where the order parameter space is the direct product between two groups $A$ and $B$: 
(a) Neither symmetry is broken, 
(b) either $A$ or $B$ is broken but the other is not broken, 
and (c) both $A$ and $B$ are broken. 
}
\end{figure}

As shown before, the phase transition nature in three-dimensional systems differs from that in two-dimensional systems even when individual order parameter spaces are the same. 
Here let us review the phase transition behavior in three-dimensional systems where the order parameter space is described by the direct product between two groups. 
Before we show some examples that have already been reported in a number of specific models, we consider generally what happens in systems where the order parameter space is described by the direct product between two groups $A \times B$. 
In these systems there are the following possible scenarios of whether two symmetries $A$ and $B$ are broken at finite temperature, which are summarized in Fig.~\ref{fig:phasetransition_concept}:
(a) Neither symmetry is broken, 
(b) either $A$ or $B$ is broken but the other is not broken,
and (c) both $A$ and $B$ are broken.
In three-dimensional systems, since a breaking of continuous symmetry at finite temperature is not prohibited in contrast to two-dimensional systems, the most possible scenario is that in Fig.~\ref{fig:phasetransition_concept} (c).
In the case of Fig.~\ref{fig:phasetransition_concept} (c), two scenarios can be considered:
(i) Two symmetries $A$ and $B$ are broken simultaneously and
(ii) $A$ and $B$ are broken successively.
An example of case (i) is the phase transition behavior in the antiferromagnetic $XY$ model. 
The order parameter space is U(1)$\times Z_2$. 
Contradictory results were reported as for the Heisenberg model on a stacked triangular lattice as mentioned above. 
Reference~\cite{Kawamura-1992} 
reported that a second-order phase transition occurs at finite temperature. 
However, the authors in Ref.~\cite{Ngo-2008a} 
concluded that a first-order phase transition occurs.
In either case, a phase transition  occurs only once in the model.
Another example is a first-order phase transition in the antiferromagnetic Heisenberg model on a face-centered-cubic lattice \cite{Diep-1989}.
The order parameter space of the model is SO(3)$\times Z_3$.
Moreover, in many cases, a phase transition occurs only once in systems with the order parameter space described by the direct product between two groups when two symmetries break at the phase transition temperature \cite{Kawamura-1992,Ngo-2008a,Diep-1989,Loison-inBook-Diep,Diep-1992,Loison-1994,Murtazeav-2010,Murtazeav-2011,Ramazanov-2011,Ngo-2011}.
Next we show an example of (ii) where the successive phase transitions occur. 
The rich phase diagram of the Bose-Hubbard model has been investigated by many kinds of methods \cite{Batrouni-1995,Boninsegni-2005,Wessel-2005,Yamamoto-2009,Danshita-2010,Kuklov-2011,Ohgoe-2012}.
At a certain parameter region, the ordered phase is the supersolid phase in which the U(1) phase symmetry and a symmetry $X$ defined by a commensurate wave vector are broken. 
Then the order parameter space is U(1)$\times X$ in the parameter region. 
In the parameter region except for the tricritical point, successive phase transitions were observed \cite{Yamamoto-2009,Kuklov-2011,Ohgoe-2012}. 
Furthermore, which phase transition occurs at higher temperature depends on the parameter. 
Recently, successive phase transitions that relate to two symmetries were also found in the site-random Heisenberg model on a three-dimensional lattice \cite{Tamura-2011a}.
As just described, a variety of phase transition natures appears in three-dimensional systems having the order parameter space $A \times B$.

The purposes of this paper are to determine the phase transition nature of the $J_1$-$J_3$ model on a stacked triangular lattice and to investigate an interlayer-interaction effect for the phase transition behavior. 
The order parameter space of the model is SO(3)$\times C_3$ for a certain parameter region, whereas the order parameter space is not the direct product between two groups for other region. 
Here we focus on the case that the order parameter space is described by SO(3)$\times C_3$.
As mentioned above, a first-order phase transition with the threefold symmetry breaking occurs in the $J_1$-$J_3$ Heisenberg model on a two-dimensional triangular lattice when the order parameter space is SO(3)$\times C_3$ \cite{Tamura-2008,Tamura-2011,Okubo-2012}.
We consider the interlayer-interaction $J_\perp$ dependence of the phase transition nature, e.g., transition temperature and latent heat.

The rest of the paper is organized as follows.
In Sec.~\ref{sec:model}, we introduce the $J_1$-$J_3$ model on a stacked triangular lattice and consider the ground state of the model.
The model consists of three types of interactions (see Fig.~\ref{fig:lattice}): ferromagnetic nearest-neighbor interaction $J_1$ and antiferromagnetic third nearest-neighbor interaction $J_3$ in each triangular layer and ferromagnetic interlayer interaction $J_\perp$.
The intralayer interactions $J_1$ and $J_3$ cause frustration.
The ground state depends on the interaction ratio $J_3/J_1$ regardless of $J_\perp$.
In Sec.~\ref{sec:three-dimension}, we show finite-temperature properties of the $J_1$-$J_3$ model on a stacked triangular lattice for $J_3/J_1= -0.85355 \cdots$ and $J_\perp/J_1=2$ by Monte Carlo simulations.
In this case, the order parameter space is SO(3)$\times C_3$.
We find that the system exhibits a first-order phase transition with breaking of the $C_3$ lattice rotational symmetry and the SO(3) symmetry of spin at finite temperature.
In Sec.~\ref{sec:quasi two-dimension}, we investigate the $J_\perp$ dependence of the phase transition behavior.
We find that as $J_\perp$ increases, setting $J_3/J_1 = -0.85355\cdots$, which is used in Sec.~\ref{sec:three-dimension}, the transition temperature increases but the latent heat decreases.
This fact is opposite to the behavior observed in typical unfrustrated three-dimensional systems such as the ferromagnetic Potts model \cite{Wu-1982} and the ferromagnetic Ising-O(3) model \cite{Kamiya-2011}.
Section~\ref{sec:conclusion} is devoted to a discussion and conclusion.
In the Appendix, we obtain the Curie-Weiss temperature from the magnetic susceptibility.


\section{Model and Ground State} \label{sec:model}

\begin{figure*}[t]
\begin{center}
\includegraphics[scale=1]{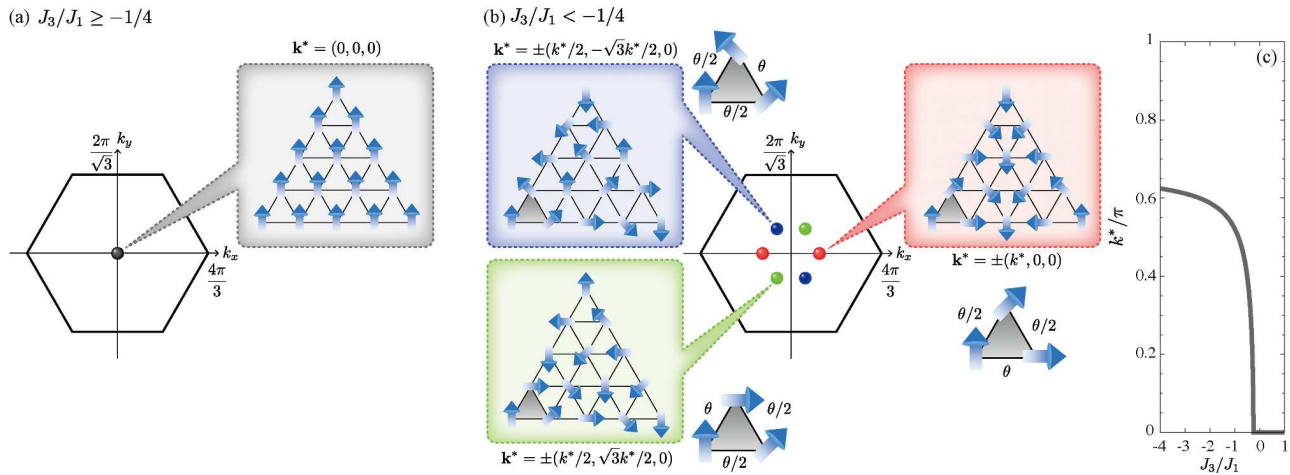} 
\end{center}
\caption{\label{fig:configuration}
(Color online) 
Explanation of ground-state properties when the nearest-neighbor interaction $J_1$ is ferromagnetic. 
(a) Position of $\mathbf{k}^*$, which minimizes the Fourier transform of interactions in the wave-vector space for $J_3/J_1\ge -1/4$.
The hexagon represents the first Brillouin zone.
A schematic of a ferromagnetic spin configuration in each triangular layer is shown.
(b) Position of $\mathbf{k}^*$ and the corresponding schematic of spiral-spin configurations in each triangular layer when $J_3/J_1 < -1/4$.
The spin configurations are depicted for $J_3/J_1= -0.85355 \cdots$ corresponding to $k^*=\pi/2$ and then $\theta=90^\circ$.
(c) The $J_3/J_1$-dependence of $k^*$.
}
\end{figure*}

We study the physical properties of a classical Heisenberg model on a stacked triangular lattice with nearest-neighbor and third-nearest-neighbor interactions.
The Hamiltonian of the system is given by
\begin{align}
\mathcal{H} = - J_1 \sum_{\langle i,j \rangle_1} \mathbf {s}_i \cdot \mathbf{s}_j
- J_3 \sum_{\langle i,j \rangle_3} \mathbf{s}_i \cdot \mathbf{s}_j
- J_\perp \sum_{\langle i,j \rangle_\perp} \mathbf{s}_i \cdot \mathbf{s}_j, \label{eq:model}
\end{align}
where ${\mathbf s}_i$ is the three-component vector spin of unit length.
The first and second sums are over all pairs of nearest-neighbor sites and that of third-nearest-neighbor sites in each triangular layer (see Fig.~\ref{fig:lattice}).
The third term represents the nearest-neighbor interlayer interactions.
Here it should be noted that the internal energy for $J_\perp >0$ is the same as that for $-J_\perp$ by applying the local gauge transformation ${\bf s}_i \to - {\bf s}_i$ for all spins in even-numbered layers. 
Then, in order to consider the phase transition nature, it is enough to study the case of the ferromagnetic interlayer interaction ($J_\perp > 0$).
Let $N=L_x \times L_y \times L_z$ be the number of spins (see Fig.~\ref{fig:lattice}).
In this paper, we study the case that $L_x=L_y=L_z=L$.

We consider the ground-state spin configuration depending on the interactions $J_1$ and $J_3$.
In general, the ground state of the Heisenberg model is a spiral-spin configuration \cite{Yoshimori-1959,Nagamiya-1967} given by
\begin{align}
\mathbf{s}_i = \mathbf{R} \cos (\mathbf{k}^* \cdot \mathbf{r}_i) - \mathbf{I} \sin (\mathbf{k}^* \cdot \mathbf{r}_i),
\end{align}
where $\mathbf{R}$ and $\mathbf{I}$ are two arbitrary orthogonal unit vectors, and $\mathbf{r}_i$ is the position vector of $i$ th site.
The vector ${\mathbf k}^*=(k_x^*,k_y^*,k_z^*)$ minimizes the Fourier transform of interactions $J(\mathbf{k})$ given by
\begin{align}
&J (\mathbf{k})/N = -J_1 \cos (k_x) - 2 J_1 \cos \left(\frac{1}{2} k_x \right) \cos \left( \frac{\sqrt{3}}{2} k_y \right) \nonumber \\
&\ \ \ - J_3 \cos (2 k_x) - 2J_3 \cos (k_x) \cos (\sqrt{3} k_y) - J_\perp \cos (k_z).
\end{align}
Here the lattice constant is set to unity.
Since we now consider the case of $J_\perp > 0$, the value of $k_z^*$ is always $0$.
In contrast, when the interlayer interaction is antiferromagnetic, $k_z^*$ is always $\pi$.
Note that the spin configuration represented by ${\bf k}$ is the same as that represented by $-{\bf k}$ in the Heisenberg model.
Here $\mathbf{k}^*$ depends on both the signs of interactions and the ratio of interactions $J_3/J_1$.

We first consider the case that $J_1$ is a ferromagnetic interaction ($J_1>0$).
When $J_3/J_1\ge -1/4$, the ground state is the ferromagnetic state, i.e., $\mathbf{k}^*=(0,0,0)$, depicted in Fig.~\ref{fig:configuration} (a). 
Then the order parameter space is $S_2$.
Thus a phase transition occurs and its universality class is expected to be the same as for the three-dimensional ferromagnetic Heisenberg model.
In contrast, when $J_3/J_1 < -1/4$, 
the ground state is a spiral-spin structure represented by one of six wave vectors
\begin{align}
\mathbf{k}^* = \pm (k^*,0,0), \ \pm (k^*/2,\sqrt{3} k^*/2,0), \ \pm (k^*/2,-\sqrt{3} k^*/2,0), \label{eq:k_stars}
\end{align}
which are depicted in Fig.~\ref{fig:configuration}(b).
The value of $k^*=|\mathbf{k}^*|$ changes between $0$ and $2\pi/3$ following the relation
\begin{align}
J_3/J_1 = - \frac{1}{2} \frac{\sin k^* + \sin \frac{1}{2} k^*}{\sin k^* + \sin 2 k^*}.
\end{align}
Figure~\ref{fig:configuration}  (c) shows a $J_3/J_1$-dependence with a value of $k^*$.
The relation means that the relative angle $\theta$ between nearest-neighbor spin pairs along one axis is $180 k^*/\pi^\circ$
and that along the other axes is $\theta/2^\circ$.
Since the system has the $120^\circ$ lattice rotational symmetry of triangular lattice $C_3$, there are three ways select the axis where the relative angle between nearest-neighbor spin pairs differs from the others, as represented by Eq.~(\ref{eq:k_stars}).
Thus, the order parameter space for $J_3/J_1 < -1/4$ is SO(3)$\times C_3$.
Next we consider the case that $J_1$ is an antiferromagnetic interaction ($J_1<0$).
When $J_3/J_1 > -1/9$, the ground state is a $120^\circ$ structure, i.e., $k^*=4\pi/3$ in Eq.~(\ref{eq:k_stars}).
Then the order parameter space is SO(3), which is the same as the order parameter space of the Heisenberg model on a stacked triangular lattice with only an antiferromagnetic nearest-neighbor interaction \cite{Kawamura-1985,Kawamura-1992,Kawamura-1998,Pelissetto-2001,Calabrese-2004,Murtazaev-2007,Zumbach-1993,Tissier-2000,Zelli-2007,Ngo-2008}.
In contrast, when $J_3/J_1 \le - 1/9$, there are degenerate ground states. 
One of the degenerate ground states is described by ${\bf k}^*=(0,2\pi/\sqrt{3},0)$. 
Degenerate ground states can be generated by applying three-dimensional rotations to the spin structure represented by ${\bf k}^*=(0,2\pi/\sqrt{3},0)$ in an appropriate way shown in Fig.~2 in Ref.~\cite{Jolicoeur-1990}.
Then the order parameter space is not well defined in this case.

Our purpose is to investigate the phase transition behavior when the order parameter space is described by the direct product between two groups.
Hereafter we focus on the parameter region $J_3/J_1 < - 1/4$ in the case of ferromagnetic $J_1$.
Throughout the paper, we use the interaction ratio $J_3/J_1=-0.85355\cdots$ so that the ground state is represented by $k^*=\pi/2$ in Eq.~(\ref{eq:k_stars}).
In this case, along one of three axes, the relative angle between nearest-neighbor spin pairs is $90^\circ$, while along the other axes, the relative angle is $45^\circ$ in the ground-state spin configuration [see Fig.~\ref{fig:configuration} (b)].
When the period of the lattice is set to $8$, the commensurate spiral-spin configuration appears in the ground state.
Then, in order to avoid the incompatibility due to the boundary effect, the linear dimension $L=8n$ ($n \in {\cal N}$) is used and the periodic boundary conditions in all directions are imposed.


\section{Finite-Temperature Properties of Stacked Model}  \label{sec:three-dimension}

\begin{figure}
\begin{center}
\includegraphics[scale=1]{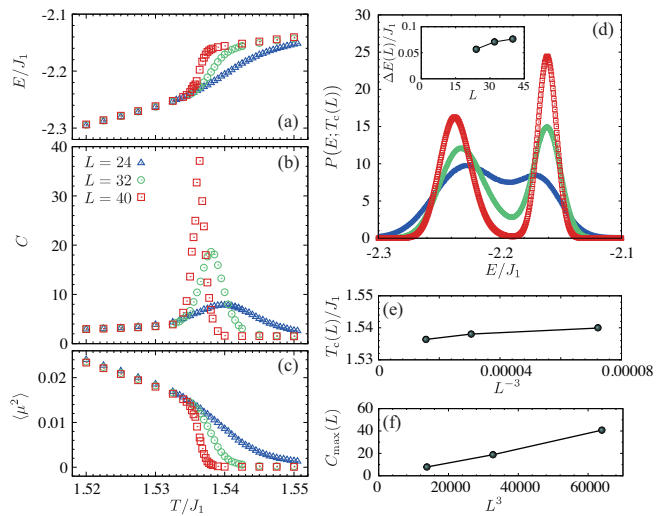} 
\end{center}
\caption{\label{fig:Ldep_Jperp2}
(Color online) 
Temperature dependence of (a) internal energy per site $E/J_1$,
(b) specific heat $C$, and (c) order parameter $\langle | {\boldsymbol \mu} |^2 \rangle$, which can detect the $C_3$ symmetry breaking of the model with $J_3/J_1= -0.85355 \cdots$ and $J_\perp/J_1= 2$ for $L=24,32,40$.
(d) Probability distribution of the internal energy $P(E;T_{\rm c}(L))$.
The inset shows the lattice-size dependence of width between bimodal peaks $\Delta E(L)/J_1$.
(e) Plot of $T_{\rm c} (L)/J_1$ as a function of $L^{-3}$.
(f) Plot of $C_{\rm max} (L)$ as a function of $L^3$.
Lines are just visual guides and error bars in all figures are omitted for clarity since their sizes are smaller than the symbol size.
}
\end{figure}

In this section, we investigate the finite-temperature properties of the Heisenberg model on a stacked triangular lattice with competing interactions given by Eq.~(\ref{eq:model}) with $J_3/J_1=-0.85355 \cdots$ and $J_\perp/J_1=2$.
Using Monte Carlo simulations with single-spin-flip heat-bath method and the over relaxation method \cite{Creutz-1987,Kanki-2005}, 
we calculate the temperature dependence of physical quantities.
Figures~\ref{fig:Ldep_Jperp2} (a) and \ref{fig:Ldep_Jperp2} (b) show the internal energy per site $E$ and specific heat $C$ for $L=24, 32, 40$.
The specific heat at temperature $T$ is given by
\begin{align}
C= N \frac{\langle E^2 \rangle - \langle E \rangle^2}{T^2},
\end{align}
where $\langle \mathcal{O} \rangle$ denotes the equilibrium value of the physical quantity $\mathcal{O}$.
Here the Boltzmann constant is set to unity.
As the system size increases, a sudden change in the internal energy is observed at a certain temperature.
In addition, the specific heat has a divergent single peak at the temperature.
These behaviors indicate the existence of a finite-temperature phase transition.
As will be shown in Sec.~\ref{sec:quasi two-dimension}, the uniform magnetic susceptibility can be used as an indicator of the phase transition.
To investigate the way of ordering, the temperature dependence of an order parameter is considered.
The order parameter ${\boldsymbol \mu}$ that can detect the $C_3$ symmetry breaking is defined by
\begin{align}
{\boldsymbol \mu} &:= \varepsilon_1 \mathbf{e}_1 + \varepsilon_2 \mathbf{e}_2 + \varepsilon_3 \mathbf{e}_3, \\
\varepsilon_\eta &:= \frac{1}{N} \sum_{\langle i,j \rangle_1 \parallel {\rm axis} \ \eta} \mathbf {s}_i \cdot \mathbf{s}_j,
\end{align}
where the subscript $\eta$ ($\eta=1,2,3$) assigns the axis (see Fig.~\ref{fig:lattice}).
The vectors ${\bf e}_\eta$ are unit vectors along the axis $\eta$ in each triangular layer, i.e., $\mathbf{e}_1=(1,0)$, $\mathbf{e}_2=(-1/2,\sqrt{3}/2)$, and $\mathbf{e}_3=(-1/2,-\sqrt{3}/2)$.
The temperature dependence of $\langle | {\boldsymbol \mu} |^2 \rangle$ is shown in Fig.~\ref{fig:Ldep_Jperp2} (c).
The order parameter abruptly increases around the temperature at which the specific heat has a divergent peak.
These results conclude that the phase transition is accompanied by the $C_3$ symmetry breaking.

To decide the order of the phase transition, we calculate the probability distribution of the internal energy at $T$, $P(E;T)=D(E)\exp(-NE/T)$, where $D(E)$ is the density of states.
When a system exhibits a first-order phase transition, the energy distribution $P(E;T)$ should be a bimodal structure at temperature $T_{\rm c}(L)$ for system size $L$.
Here $T_{\rm c}(L)$ is the temperature at which the specific heat becomes the maximum value $C_{\rm max}(L)$.
To obtain $T_{\rm c}(L)$ and $C_{\rm max}(L)$, we perform the reweighting method \cite{Ferrenberg-1988}.
Figure~\ref{fig:Ldep_Jperp2} (d) shows $P(E;T_{\rm c}(L))$ for system sizes $L=24, 32, 40$.
As stated above, the bimodal structure in the energy distribution suggests a first-order phase transition.

To confirm whether the first-order phase transition behavior remains in the thermodynamic limit, we perform two types of analysis.
One is the finite-size scaling and the other is a naive analysis of probability distribution $P(E;T_{\rm c}(L))$.
The scaling relations for the first-order phase transition in $d$-dimensional systems \cite{Challa-1986} are given by
\begin{align}
T_{\rm c} (L) &= a L^{-d} + T_{\rm c}, \label{scaling1}\\
C_{\rm max} (L) &\propto \frac{(\Delta E)^2 L^d}{4 T_{\rm c}^2}, \label{scaling2} 
\end{align}
where $T_{\rm c}$ and $\Delta E$ are, respectively, the transition temperature and the latent heat in the thermodynamic limit.
The coefficient of the first term in Eq.~(\ref{scaling1}), $a$, is a constant.
Figures~\ref{fig:Ldep_Jperp2} (e) and \ref{fig:Ldep_Jperp2} (f) show the scaling plots for $T_{\rm c} (L)/J_1$ and $C_{\rm max} (L)$, respectively.
Figure~\ref{fig:Ldep_Jperp2} (e) indicates that $T_{\rm c}$ is a nonzero value in the thermodynamic limit.
Figure~\ref{fig:Ldep_Jperp2} (f) shows an almost linear dependence of $C_{\rm max}(L)$ as a function of $L^3$.
However, using the finite-size scaling, we cannot obtain the transition temperature and latent heat in the thermodynamic limit with high accuracy because of the strong finite-size effect.
Next, we directly calculate the size dependence of the width between bimodal peaks of the energy distribution shown in Fig.~\ref{fig:Ldep_Jperp2} (d).
The width for the system size $L$ is represented by $\Delta E(L)=E_+(L)-E_-(L)$ where $E_+(L)$ and $E_-(L)$ are the averages of the Gaussian function in the high-temperature phase and that in the low-temperature phase, respectively.
In the thermodynamic limit, each Gaussian function becomes the delta-function and then $\Delta E(L)$ converges to $\Delta E$ \cite{Challa-1986}.
The inset of Fig.~\ref{fig:Ldep_Jperp2} (d) shows the size dependence of the width $\Delta E(L)/J_1$.
The width enlarges as the system size increases, which indicates that the latent heat is a nonzero value in the thermodynamic limit.
The results shown in Fig.~\ref{fig:Ldep_Jperp2} conclude that the model given by Eq.~(\ref{eq:model}) exhibits the first-order phase transition with the $C_3$ symmetry breaking at finite temperature.

\begin{figure}
\begin{center}
\includegraphics[scale=1]{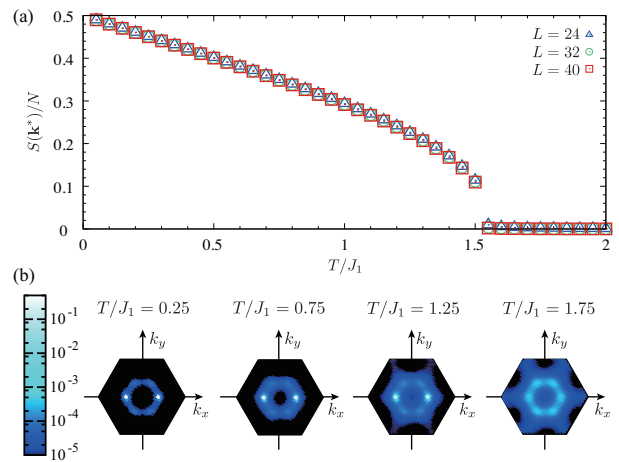} 
\end{center}
\caption{\label{fig:sq}
(Color online) 
(a) Temperature dependence of the largest value of structure factors $S(\mathbf{k}^*)$ calculated by six wave vectors in Eq.~(\ref{eq:k_stars}) for $J_3/J_1=-0.85355\cdots$ and $J_\perp/J_1=2$.
Error bars are omitted for clarity since their sizes are smaller than the symbol size.
(b) Structure factors at $k_z=0$ in the first Brillouin zone at several temperatures for $L=40$.
}
\end{figure}

We further investigate the way of spin ordering.
As mentioned above, the order parameter space of the system is SO(3)$\times C_3$.
It was confirmed that the $C_3$ symmetry breaks at the first-order phase transition point.
In the antiferromagnetic Heisenberg model on a stacked triangular lattice with only a nearest-neighbor interaction where the order parameter space is SO(3), 
a single peak is observed for the temperature dependence of the specific heat \cite{Kawamura-1985,Kawamura-1992}.
The peak indicates the finite-temperature phase transition between the paramagnetic state and magnetic ordered state where the SO(3) symmetry is broken.
Then, in our model, the SO(3) symmetry should break at the first-order phase transition point since the specific heat has a single peak corresponding to the first-order phase transition.
To confirm this, we calculate the temperature dependence of the structure factor of spin:
\begin{align}
S (\mathbf{k}) := \frac{1}{N} \sum_{i,j} \langle \mathbf{s}_i \cdot \mathbf{s}_j \rangle {\rm e}^{- i \mathbf{k} \cdot (\mathbf{r}_i-\mathbf{r}_j)},
\end{align}
which is the magnetic order parameter for spiral-spin states.
When the magnetic ordered state described by $\mathbf{k}^*$ where the SO(3) symmetry is broken appears,
$S(\mathbf{k}^*)$ becomes a finite value in the thermodynamic limit.
Figure~\ref{fig:sq} (a) shows the temperature dependence of the largest value of structure factors $S(\mathbf{k}^*)$ calculated by six wave vectors in Eq.~(\ref{eq:k_stars}).
Here $S(\mathbf{k}^*)$ becomes zero in the thermodynamic limit above the first-order phase transition temperature.
The structure factor $S(\mathbf{k}^*)$ becomes a nonzero value at the first-order phase transition temperature.
Moreover, as temperature decreases, the structure factor $S({\bf k}^*)$ increases. 
The structure factors at $k_z=0$ in the first Brillouin zone at several temperatures for $L=40$ are also shown in Fig.~\ref{fig:sq} (b).
As mentioned in Sec.~\ref{sec:model}, the spiral-spin structure represented by ${\bf k}$ is the same as that represented by $-{\bf k}$ in the Heisenberg models.
Figure~\ref{fig:sq} (b) confirms that one distinct wave vector is chosen from three types of ordered vectors below the first-order phase transition point, which is further evidence of the $C_3$ symmetry breaking at the first-order phase transition temperature.

Before we end this section, let us mention a phase transition nature in the $J_1$-$J_2$ Heisenberg model with interlayer interaction $J_\perp$ on a stacked triangular lattice. 
In Refs.~\cite{Loison-1994,Murtazeav-2010,Murtazeav-2011,Ramazanov-2011},
the authors studied the phase transition behavior of the model when $J_1$ and $J_2$ are antiferromagnetic interactions. 
For large $J_2/J_1$, a phase transition between the paramagnetic phase and ordered incommensurate spiral-spin structure phase occurs at finite temperature. 
In the parameter region, the order parameter space is SO(3)$\times C_3$ and a second-order phase transition with threefold symmetry occurs \cite{Loison-1994}, which differs from the result obtained in this section. 
However, in frustrated spin systems, a different phase transition nature happens even when the symmetry that is broken at the phase transition temperature is the same as for other models. 
For example, in the $J_1$-$J_3$ Heisenberg model on a triangular lattice, a first-order phase transition with threefold symmetry breaking occurs when $J_3/J_1 < -1/4$ and $J_1>0$. 
It is well known that the simplest model that exhibits a phase transition with threefold symmetry breaking is the three-state ferromagnetic Potts model \cite{Wu-1982}.
The three-state ferromagnetic Potts model in two dimensions exhibits a second-order phase transition. 
It is no wonder that our obtained result differs from the results in the previous study \cite{Loison-1994}.


\section{Dependence on Interlayer interaction}  \label{sec:quasi two-dimension}

\begin{figure}
\begin{center}
\includegraphics[scale=1]{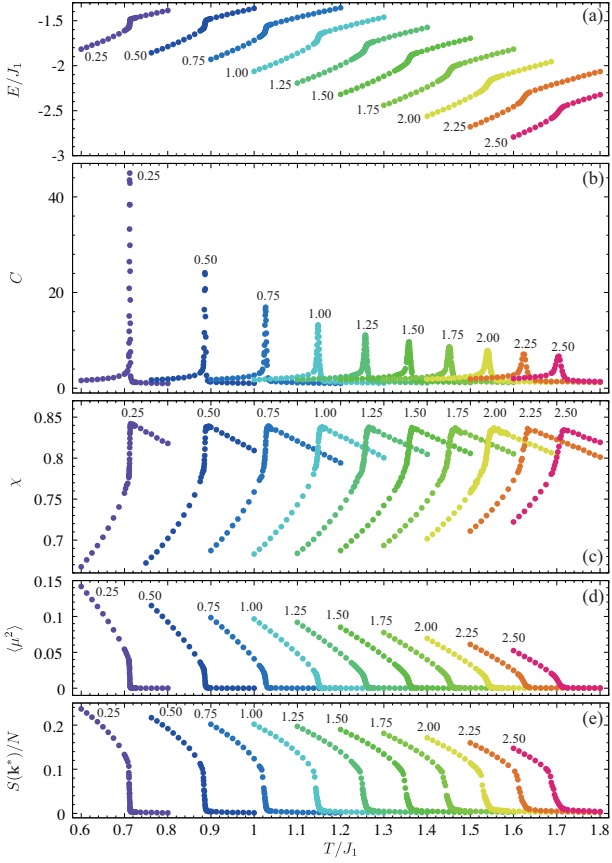} 
\end{center}
\caption{\label{fig:Jpdep_temp}
(Color online) 
Interlayer-interaction $J_\perp/J_1$ dependence of (a) internal energy per site $E/J_1$, 
(b) specific heat $C$, (c) uniform magnetic susceptibility $\chi$,
(d) order parameter $\langle | {\boldsymbol \mu} |^2 \rangle$, which can detect the $C_3$ symmetry breaking,
and (e) largest value of structure factors $S(\mathbf{k}^*)$ calculated by six wave vectors in Eq.~(\ref{eq:k_stars}) for $L=24$.
Error bars in all figures are omitted for clarity since their sizes are smaller than the symbol size.
}
\end{figure}

\begin{figure}
\begin{center}
\includegraphics[scale=1]{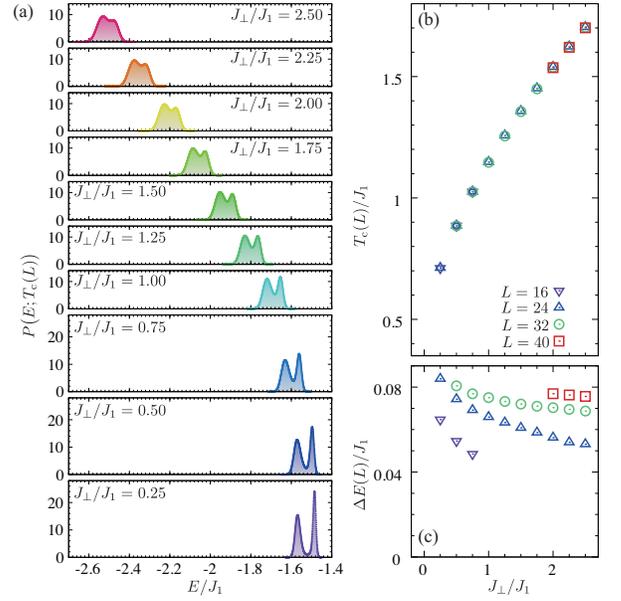} 
\end{center}
\caption{\label{fig:Jpdep_hist}
(Color online) 
(a) Interlayer-interaction $J_\perp/J_1$ dependence of the probability distribution of internal energy $P(E;T_{\rm c} (L))$ when the specific heat becomes the maximum value for $L=24$.
(b) The $J_\perp/J_1$ dependence of $T_{\rm c} (L)/J_1$ at which the specific heat becomes the maximum value for $L=16-40$.
(c) The $J_\perp/J_1$ dependence of the width between bimodal peaks of the energy distribution $\Delta E (L)/J_1$.
Error bars in all figures are omitted for clarity since their sizes are smaller than the symbol size.
}
\end{figure}

In this section, we study interlayer-interaction dependence of the phase transition behavior.
Here we set the interaction ratio $J_3/J_1=-0.85355\cdots$ at which the ground state is represented by $k^*=\pi/2$ in Eq.~(\ref{eq:k_stars}), as in the previous section.
In the previous section, we considered the case that $J_\perp/J_1=2$.
We found that the first-order phase transition with the $C_3$ symmetry breaking occurs 
and breaking of the SO(3) symmetry at the first-order phase transition point was confirmed.

Figure~\ref{fig:Jpdep_temp} shows the temperature dependence of physical quantities for $L=24$ with several interlayer interactions $0.25 \le J_\perp/J_1 \le 2.5$, setting $J_3/J_1=-0.85355\cdots$.
Figure~\ref{fig:Jpdep_temp} (a) shows the internal energy as a function of temperature, which displays that the temperature at which the sudden change of the internal energy appears increases as $J_\perp/J_1$ increases.
In other words, Fig.~\ref{fig:Jpdep_temp} (a) indicates that the first-order phase transition temperature monotonically increases as a function of $J_\perp/J_1$.
In addition, the energy difference between the high-temperature phase and low-temperature phase decreases as $J_\perp/J_1$ increases.
These behaviors are supported by the temperature dependence of the specific heat shown in Fig.~\ref{fig:Jpdep_temp} (b).
Furthermore,
in the specific heat,
no peaks, except the first-order phase transition temperature, are observed by changing the value of $J_\perp/J_1$.
Figure~\ref{fig:Jpdep_temp} (c) shows the uniform magnetic susceptibility $\chi$, which is calculated by
\begin{align}
\chi = \frac{N J_1 \langle |{\mathbf m}|^2 \rangle}{T},
\quad
{\mathbf m} = \frac{1}{N} \sum_i {\mathbf s}_i,
\end{align}
where ${\mathbf m}$ is the uniform magnetization.
The uniform magnetic susceptibility has the sudden change at the first-order phase transition temperature.
As stated in Sec.~\ref{sec:three-dimension}, it can be used as an indicator of the first-order phase transition. 
Note that the magnetic susceptibility of the model with $J_\perp$ differs from that with $-J_\perp$.
However, the sudden change in $\chi$ at the first-order phase transition temperature is also observed when the interlayer interaction is antiferromagnetic.
We obtain the Curie-Weiss temperature from the magnetic susceptibility for several $J_\perp/J_1$, including the case of antiferromagnetic $J_\perp$, which will be shown in the Appendix.
In addition, Figs.~\ref{fig:Jpdep_temp} (d) and \ref{fig:Jpdep_temp} (e) confirm that phase transitions always accompany the $C_3$ lattice rotational symmetry breaking and breaking of the global rotational symmetry of spin, the SO(3) symmetry, for the considered $J_\perp/J_1$, respectively.

Next, in order to consider the $J_\perp/J_1$ dependence of the latent heat,
we calculate the probability distribution of the internal energy $P(E;T_{\rm c} (L))$ for several values of $J_\perp/J_1$ shown in Fig.~\ref{fig:Jpdep_hist} (a).
The width between bimodal peaks decreases as $J_\perp/J_1$ increases.
Furthermore, we calculate interlayer-interaction dependences of $T_{\rm c} (L)/J_1$ and $\Delta E (L)/J_1$ for $L=16$-$40$ which are shown in Figs.~\ref{fig:Jpdep_hist} (b) and \ref{fig:Jpdep_hist} (c).
As $J_\perp/J_1$ increases, $T_{\rm c} (L)/J_1$ monotonically increases and $\Delta E (L)/J_1$ decreases for each system size.
In addition, $\Delta E(L)/J_1$ increases as the system size increases.
Here $\Delta E(L)/J_1$ in the thermodynamic limit corresponds to the latent heat.
Thus Fig.~\ref{fig:Jpdep_hist} (c) suggests that the latent heat decreases as $J_\perp/J_1$ increases in the thermodynamic limit.


\section{Discussion and Conclusion}  \label{sec:conclusion}

In this paper, the nature of the phase transition of the Heisenberg model on a stacked triangular lattice was studied by Monte Carlo simulations.
In our model, there are three kinds of interactions: the ferromagnetic nearest-neighbor interaction $J_1$ and antiferromagnetic third nearest-neighbor interaction $J_3$ in each triangular layer and the ferromagnetic nearest-neighbor interlayer interaction $J_\perp$.
When  $J_3/J_1<-1/4$, the ground state is a spiral-spin structure in which the $C_3$ symmetry is broken as in the case of two-dimensional $J_1$-$J_3$ Heisenberg model on a triangular lattice \cite{Tamura-2008,Tamura-2011}.
Then the order parameter space in the case is described by SO(3)$\times C_3$.

In Sec.~\ref{sec:three-dimension}, we studied the finite-temperature properties of the system with $J_3/J_1=-0.85355\cdots$ and $J_\perp/J_1=2$.
We found that a first-order phase transition takes place at finite temperature.
The temperature dependence of the order parameter indicates that the $C_3$ symmetry breaks at the transition temperature, which is the same feature as in the two-dimensional case \cite{Tamura-2008,Tamura-2011}.
We also calculated the temperature dependence of the structure factor at the wave vector representing the ground state, 
which is the magnetic order parameter for spiral-spin states.
The result shows that the SO(3) symmetry breaks at the transition temperature.

In Sec.~\ref{sec:quasi two-dimension}, we investigated the interlayer interaction effect on the nature of phase transitions.
We confirmed that the first-order phase transition occurs for $0.25 \le J_\perp/J_1 \le 2.5$ and $J_3/J_1=-0.85355\cdots$ which was used in Sec.~\ref{sec:three-dimension}.
We could not determine the existence of the first-order phase transition for $J_\perp/J_1< 0.25$ or $J_\perp/J_1 > 2.5$ by Monte Carlo simulations.
In the parameter ranges, the width of two peaks in the probability distribution of the internal energy cannot be estimated easily because of the finite-size effect.
It is a remaining problem to determine whether a second-order phase transition occurs for large $J_\perp/J_1$ as in the $J_1$-$J_2$ Heisenberg model on a stacked triangular lattice \cite{Loison-1994}.
As the ratio $J_\perp/J_1$ increases, the first-order phase transition temperature monotonically increases but the latent heat decreases.
This is opposite to the behavior observed in typical unfrustrated three-dimensional systems that exhibit a first-order phase transition at finite temperature.
For example, the $q$-state Potts model with ferromagnetic intralayer and interlayer interactions ($q\ge 3$) is a fundamental model that exhibits a temperature-induced first-order phase transition with $q$-fold symmetry breaking \cite{Wu-1982}.
From a mean-field analysis of the ferromagnetic Potts model \cite{Kihara-1954,Wu-1982}, 
as the interlayer interaction increases, both the transition temperature and the latent heat increase.
The same behavior was observed in the Ising-O(3) model on a stacked square lattice \cite{Kamiya-2011}.
As just described, in general, if the parameter that can stabilize the ground state becomes large, the transition temperature increases and the latent heat increases \cite{Kihara-1954,Wu-1982,Kamiya-2011}.
Furthermore, in conventional systems, both the transition temperature and the latent heat are expressed by the value of an effective interaction obtained by a characteristic temperature such as the Curie-Weiss temperature.
However, in our model, the Curie-Weiss temperature does not characterize the first-order phase transition, as will be shown in the Appendix.
Thus our result is an unusual behavior.
The investigation of the essence of the obtained results is a remaining problem.


\section*{Acknowledgment}
R.T. was partially supported by a Grand-in-Aid for Scientific Research (C) (Grant No. 25420698) and National Institute for Materials Science.
S.T. was partially supported by a Grand-in-Aid for JSPS Fellows (Grant No 23-7601).
The computations in the present work were performed on computers at the Supercomputer Center, Institute for Solid State Physics, University of Tokyo. 



\appendix

\section{Interlayer-interaction dependence of the Curie-Weiss Temperature}
\label{sec:cwtemp}

\begin{figure}[t]
\begin{center}
\includegraphics[scale=1]{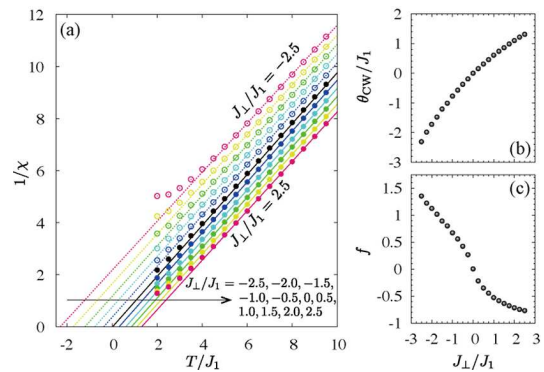} 
\end{center}
\caption{\label{fig:cwtemp}
(Color online) 
(a) Inverse of the magnetic susceptibility $\chi^{-1}$ as a function of temperature with $J_3/J_1=-0.85355\cdots$ and several $J_\perp/J_1$ for $L=24$.
The lines are obtained by the least-squares estimation.
(b) The $J_\perp/J_1$ dependence of the Curie-Weiss temperature.
(c) The $J_\perp/J_1$ dependence of the ratio $f=-\theta_{\rm CW}/T_{\rm c}$.
}
\end{figure}

In this section, we obtain the Curie-Weiss temperature for several $J_\perp/J_1$ including the case of the antiferromagnetic interlayer interaction.
Here we also use the interaction ratio $J_3/J_1=-0.85355\cdots$, which was used in Secs.~\ref{sec:three-dimension} and \ref{sec:quasi two-dimension}.
As mentioned in Sec.~\ref{sec:model}, the phase transition behavior of the model with $J_\perp$ is the same as that with $-J_\perp$, which is proved by the local gauge transformation. 
However, 
the Curie-Weiss temperature for $J_\perp$ differs from that for $-J_\perp$.
Figure~\ref{fig:cwtemp} (a) shows the inverse of the magnetic susceptibility $\chi^{-1}$ as a function of temperature in the high-temperature region for $L=24$.
In general, the temperature dependence of the magnetic susceptibility at high temperatures is expressed as
\begin{eqnarray}
 \chi = \frac{A}{T-\theta_{\rm CW}},
\end{eqnarray}
where $A$ is the Curie constant and $\theta_{\rm CW}$ is the Curie-Weiss temperature.
The Curie-Weiss temperature $\theta_{\rm CW}$ represents a characteristic temperature of magnetic systems.
The interlayer-interaction dependence of the Curie-Weiss temperature is shown in Fig.~\ref{fig:cwtemp} (b).
The Curie-Weiss temperature dependence is not symmetric about the origin.
The first-order phase transition temperature of the system with $J_\perp$ is the same as that with $-J_\perp$.
Then the transition temperature does not relate to the Curie-Weiss temperature.

Next we consider the $J_\perp/J_1$ dependence of the ratio of two characteristic temperatures $f:=-\theta_{\rm CW}/T_{\rm c}$.
In frustrated systems, $f$ is a useful quantity to express the degree of frustration and is called the frustration parameter.
Figure~\ref{fig:cwtemp} (c) depicts the interlayer-interaction dependence of $f$.
The value of $f$ is not a characteristic quantity which expresses the nature of first-order phase transition as well as the Curie-Weiss temperature.

\end{document}